\documentclass[twocolumn,aps]{revtex4}
\usepackage{natbib}
\usepackage{graphicx}

\newcommand {\omegape}	{\omega_\mathrm{pe}}
\newcommand {\omegaps}	{\omega_\mathrm{ps}}
\newcommand {\sums}	{\sum_\mathrm{s}}
\newcommand {\ns}	{n_\mathrm{s}}
\newcommand {\Vs}	{V_\mathrm{s}}
\newcommand {\aths}	{a_\mathrm{s}}
\newcommand {\athsx}	{a_{\mathrm{s}x}}
\newcommand {\athsy}	{a_{\mathrm{s}y}}
\newcommand {\athsz}	{a_{\mathrm{s}z}}
\newcommand {\kDs}	{k_\mathrm{Ds}}
\newcommand {\zetas}	{\zeta_\mathrm{s}}
\newcommand {\etas}	{\eta_\mathrm{s}}
\newcommand {\lambdae}	{\lambda_\mathrm{e}}
\newcommand {\nez}	{n_\mathrm{e0}}
\newcommand {\me}	{m_\mathrm{e}}
\newcommand {\mi}	{m_\mathrm{i}}
\newcommand {\Te}	{T_\mathrm{e}}
\newcommand {\Ti}	{T_\mathrm{i}}
\newcommand {\Tix}	{T_{\mathrm{i},x}}
\newcommand {\Tiy}	{T_{\mathrm{i},y}}
\newcommand {\Tiz}	{T_{\mathrm{i},z}}
\newcommand {\athe}	{a_\mathrm{e}}
\newcommand {\athi}	{a_\mathrm{i}}
\newcommand {\kB}	{k_\mathrm{B}}
\begin{document}

\title{Electrostatic and electromagnetic instabilities associated with electrostatic shocks:
two-dimensional particle-in-cell simulation}
\author{Tsunehiko N. Kato}
\email[E-mail: ]{kato-t@ile.osaka-u.ac.jp}
\author{Hideaki Takabe}
\affiliation{
Institute of Laser Engineering, Osaka University,
2-6 Yamada-oka, Suita, Osaka 565-0871, Japan
}

\date{\today}

\begin{abstract}
A two-dimensional electromagnetic particle-in-cell simulation
with the realistic ion-to-electron mass ratio of 1836
is carried out
to investigate the electrostatic collisionless shocks in relatively
high-speed ($\sim 3000$ km s$^{-1}$) plasma flows
and also the influence of both electrostatic and electromagnetic instabilities,
which can develop around the shocks,
on the shock dynamics.
It is shown that the electrostatic ion-ion instability can develop
in front of the shocks,
where the plasma is under counter-streaming condition,
with highly oblique wave vectors
as was shown previously. %by Karimabadi, Omidi and Quest (1991).
The electrostatic potential generated by the electrostatic ion-ion instability
propagating obliquely to the shock surface
becomes comparable with the shock potential and
finally the shock structure is destroyed.
It is also shown that in front of the shock
the beam-Weibel instability gradually grows as well,
consequently suggesting that
the magnetic field generated by the beam-Weibel instability becomes important
in long-term evolution of the shock and
the Weibel-mediated shock forms long after the electrostatic shock vanished.
It is also observed that the secondary electrostatic shock forms
in the reflected ions in front of the primary electrostatic shock.
\end{abstract}
\maketitle

%%%%%
\section{Introduction}
%%%%%
Collisionless shock is one of the most interesting
phenomena in plasma physics; it dissipates the kinetic energy
of the plasma flow into the thermal energy and the electromagnetic energy
not by the Coulomb collision
but by the collective effect associated with the electric and the magnetic fields.
For example,
the universe is filled with hot, tenuous collisionless plasmas and
a variety of collisionless shocks are produced due to violent phenomena,
such as supernova explosions.
These shocks are believed to accelerate charged particles
to high energies to generate cosmic rays.

The electrostatic shock \citep{Moiseev63}, or the ion-acoustic shock \citep{Chen},
is also one of the collisionless shocks and
it forms in unmagnetized collisionless electron-ion plasmas
if the Mach number is not so large
and the temperature ratio of electrons to ions is relatively large
\citep{Mason72}.
The electrostatic shocks were observed in various experiments
with double-plasma devices \citep{Taylor70, Ikezi73, Bailung08},
with Q-machines \citep{Takeuchi98},
with photo-ionized plasmas \citep{Cohn72},
and with laser plasmas \citep{Koopman67,Romagnani08}.
Recent experiments with intense lasers
also showed a possible formation of the electrostatic shock
at a high shock speed of $\sim 1000$ km s$^{-1}$ \citep{Morita09}.
In space, these shocks are observed, for example, in the auroral zone of the Earth \citep{Mozer81}
as well.
In the astrophysical context,
there are no clear observations of the electrostatic shocks so far.
However,
they can also be driven in the universe at a wide range of the plasma flow speed.
From the theoretical point of view, the electrostatic shock has been
investigated with %one-dimensional
hybrid or particle-in-cell (PIC) simulations extensively
\citep[e.g.,][]{Forslund70b,Mason71,Mason72}.
Recent simulations also showed
a possible formation of very high Mach number electrostatic shocks \citep{Sorasio06}.

The requirement of the relatively high electron-to-ion temperature ratio
for the electrostatic shocks is also one of the conditions for
the electrostatic ion-ion instability \citep{Stringer64,Ohnuma65,Fried66,Gary87},
which is also called the ion/ion acoustic instability,
to develop.
Since there exist the reflected ions in front of the electrostatic shocks,
this instability can grow to generate electrostatic waves  there.
The wave vector of the instability is oblique or even
almost perpendicular to the streaming direction \citep{Forslund70a,Gresillon75}
and therefore multi-dimensional simulations are necessary to investigate
the effect of the instability.
Although most of the simulations were carried out in one dimension
and therefore could not deal with the instability,
\citet{Karimabadi91} carried out a two-dimensional electrostatic hybrid simulation
and showed that
the electrostatic ion-ion instability indeed develops in front of the electrostatic shock
and affects the structure of the shock significantly.
(It should be noted that recently \citet{Ohira08} showed that
this instability can also develop in the foot region of 
the magnetized collisionless shocks in supernova remnants.)

In astrophysical plasmas and laboratory plasmas with recent laser facilities,
flows of collisionless plasma
whose velocity is even faster than 1000 km s$^{-1}$
can be generated.
In such high-speed flows,
the electromagnetic instabilities, e.g., the Weibel-type instabilities \citep{Weibel59, Fried59},
appears to be important as well as the electrostatic instabilities
and can affect the evolution of the electrostatic shocks.
The simulation by \citet{Karimabadi91} was however an electrostatic one
and the effects of such electromagnetic instabilities were therefore not included.
In addition, since the simulation box size of their simulation was not large compared with
the structures generated by the electrostatic instability,
larger scale simulations are also desirable.

In this paper, we investigate the electrostatic shocks
propagating at a relatively high speed
with a two-dimensional electromagnetic PIC simulation.
In particular,
we focus on the influence of the electrostatic ion-ion instability
and the Weibel-type instabilities on the shock formation.

%%%%%
\section{Linear analysis}
\label{sec:linear}
%%%%%
The electrostatic shocks can form in two counter-streaming collisionless plasmas
with a relatively high electron-to-ion temperature ratio
as shown by the previous simulations \citep{Forslund70b,Mason72}.
In such situations,
the plasma becomes counter-streaming condition in front of the shock
and some instabilities (electrostatic and/or electromagnetic)
can develop there to affect the shock dynamics.
In this section,
we present the linear analysis of the electrostatic ion-ion instability
and the electromagnetic beam-Weibel instability in counter-streaming plasmas
with wave vectors in arbitrary direction.

Consider a counter-streaming plasma consisting of Maxwellian beams;
each beam of the species s has
the drift velocity of $\Vs$ along the $x$-axis and the thermal velocity of $\aths$.
The non-perturbed (zeroth order) distribution function for each species is given by
\begin{equation}
	f_0^{(\mathrm{s})}(v_x, v_y, v_z)
	= \frac{\ns}{\pi^{3/2} \aths^3}
		\exp \left[
		-\frac{(v_x - \Vs)^2 + v_y^2 + v_z^2}{\aths^2} \right],
\end{equation}
where $\ns$ is the number density.
For this system,
the linear dispersion relation of the electrostatic mode is given by
\begin{equation}
	k^2 + 2\sums \kDs^2 \left( 1 + \zetas Z(\zetas) \right) = 0,
	\label{eq:disp_rel_es}
\end{equation}
where
$k$ is the magnitude of the wave vector,
$\theta$ is the angle between the wave vector and the $x$-axis,
\begin{equation}
	\kDs \equiv \omegaps / \aths,
	\qquad
	\omegaps \equiv \left( \frac{4\pi \ns q_\mathrm{s}^2}{m_\mathrm{s}} \right)^{1/2},
\end{equation}
and
\begin{equation}
	\zetas = \zetas(\omega, k, \theta) \equiv (\omega / k - \Vs\cos\theta) / \aths.
\end{equation}
The function $Z(\zeta)$ called the plasma dispersion function \citep{Fried} is defined by
\begin{equation}
	Z(\zeta) \equiv \pi^{-1/2} \int_{-\infty}^{\infty} \frac{e^{-z^2}}{z - \zeta} dz
\end{equation}
on the upper half of the complex $\zeta$-plane ($\mathrm{Im}(\zeta) > 0$)
and analytically continued into the lower half-plane.
Note that
the solution of the dispersion relation (\ref{eq:disp_rel_es}) with
$\mathrm{Im}(\omega) > 0$ means an unstable mode.

\begin{figure}[!hb]
\resizebox{8.cm}{!}{\includegraphics*{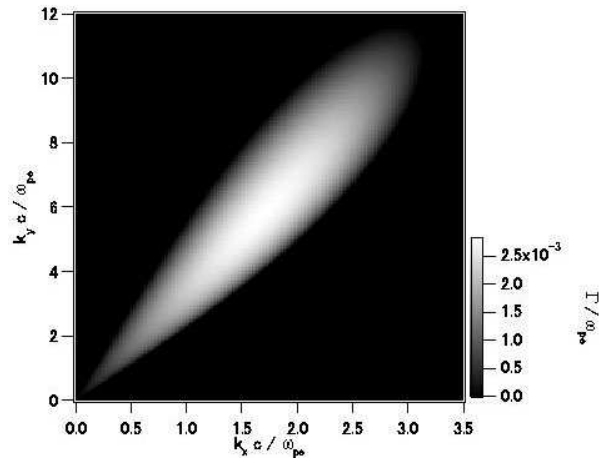}} \\[-0.6mm]
\caption{
Linear growth rate of the electrostatic ion-ion instability
for the symmetrical beam system of electron-ion plasma
with the drift velocity of $0.01c$
and the electron-to-ion temperature ratio of 9.
The horizontal and the vertical axes are
the wave vector in the $x$-direction, $k_x$,
and that in the $y$-direction, $k_y$, respectively.
The gray-scale shows the linear growth rate
normalized to the electron plasma frequency.
}
\label{fig:es_disp}
\end{figure}

Here, we consider a symmetric Maxwellian beam system
consisting of two electron beams (s = e+ and e-)
and two ion beams (s = i+ and i-)
with $n_{e+} = n_{e-} = n_{i+} = n_{i-} = \nez$,
$V_\mathrm{e+} = V_\mathrm{i+} = V$
and $V_\mathrm{e-} = V_\mathrm{i-} = -V$.
Figure \ref{fig:es_disp} shows the linear growth rate of
the electrostatic ion-ion instability \citep{Stringer64,Ohnuma65,Fried66,Gary87},
$\Gamma \equiv \mathrm{Im}(\omega)$,
numerically calculated for a symmetrical beam system
with a relatively high electron-to-ion temperature ratio.
For later convenience,
we take the following parameters:
$\mi/\me = 1836$,
$V = 0.01c \sim 3000$ km s$^{-1}$,
$\Te = 11.3$ keV ($\athe = 0.21c$),
$\Ti = 1.25$ keV ($\athi = 1.6 \times 10^{-3}c$),
and thus $\Te/\Ti = 9$,
where $c$ is the speed of light,
$\Te$ and $\Ti$ are the electron and the ion temperatures, respectively.
%%%
%%%
The maximum growth rate is obtained as
$\Gamma = 2.8 \times 10^{-3} \omegape$
at $k = 6.0 \ \omegape c^{-1}$
and $\theta = 74^\circ$,
where $\omegape \equiv (4\pi \nez e^2 / \me)^{1/2}$
is the electron plasma frequency defined for the mean electron number density $\nez$;
therefore the dominant mode is directed
almost perpendicular to the streaming direction.
The corresponding typical wavelength is given by
$\lambda = 2\pi/k = 6.7 \times 10^{-1} \lambdae$,
where $\lambdae = c \omegape^{-1}$ is the electron skin depth.

\begin{figure}[!hb]
\resizebox{8.cm}{!}{\includegraphics*{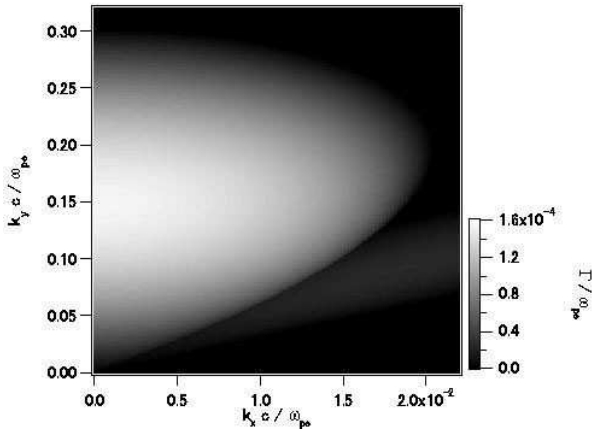}} \\[-0.6mm]
\caption{
Linear growth rate of the electromagnetic beam-Weibel instability
for the same beam system as in Fig.~\ref{fig:es_disp}.
}
\label{fig:em_disp}
\end{figure}

On the other hand,
one of the two electromagnetic modes whose electric field
lies on the plane made by the $x$-axis and the wave vector
can be unstable in this system, too;
this is the beam-Weibel instability (or the filamentation instability) \citep{Fried59},
which is distinguished from the (ordinary) Weibel instability \citep{Weibel59}
that grows in a plasma with anisotropic temperatures.
The dispersion relation of the beam-Weibel instability is given by
\begin{equation}
	\omega^2 - (kc)^2 + \sums \omegaps^2
	\left[ \frac{\aths^2 + 2 (\Vs \sin\theta)^2}{\aths^2} (1+\zetas Z(\zetas)) - 1\right]
	= 0.
\end{equation}
Figure \ref{fig:em_disp} shows the linear growth rate of this instability
for the same system as in Fig.~\ref{fig:es_disp}.
The maximum growth rate is obtained as $\Gamma = 1.5 \times 10^{-4} \omegape$
at $k = 0.15 \ \omegape c^{-1}$ (the typical wavelength of $\lambda = 4.2 \times 10^{1} \lambdae$)
and $\theta = 90^\circ$.
Although the growth rate of this beam-Weibel instability is smaller by one order of magnitude
than that of the electrostatic ion-ion instability,
the magnetic field generated by this instability
can be important during the long-term evolution.
%%%
%%%

%%%%%
\section{Simulation}
%%%%%
We perform a two-dimensional electromagnetic PIC simulation
to investigate the electrostatic shocks
together with the instabilities associated with the shock.
The simulation code used is a relativistic, electromagnetic, particle-in-cell code
with two spatial and three velocity dimensions, namely 2D3V code,
developed based on a standard method described in \citet{Birdsall};
this code was also used to demonstrate the formation of the Weibel-mediated
collisionless shocks at relativistic speed in pair plasma \citep{Kato07}
and at nonrelativistic speed in electron-ion plasma \citep{Kato08}.
Furthermore,
it was used to derive the scaling law for the model
experiment in laboratory to demonstrate the formation of the Weibel-mediated
collisionless shocks in laboratory \citep{Takabe08} as well.
Thus, our code can deal with the electromagnetic modes as well as
the (oblique) electrostatic modes.
The basic equations of the simulation are
the Maxwell's equations
and the (relativistic) equation of motion of particles.
In the following,
the simulation plane is regarded as the $x-y$ plane
and the $z$-axis is taken perpendicular to the plane.
We take $\omegape^{-1}$ as the unit of time
and the electron skin depth $\lambdae = c \omegape^{-1}$ as the unit of length
(for example, for $\nez = 1$ cm$^{-3}$, $\omegape^{-1} = 1.77 \times 10^{-5}$ s
and $\lambdae = 5.31 \times 10^{5}$ cm.)
The units of electric and magnetic fields are $E_* = B_* = c (4\pi \nez \me)^{1/2}$.

%%%
\subsection{Initial condition}
%%%
Here,
we take the following parameters,
which is similar to those of the one-dimensional simulation
shown in Fig.~1 (c) of \citet{Forslund70b}.
The ratio of the ion mass to the electron mass is 1836.
The grid size is $N_x \times N_y = 4096 \times 1024$ 
and the number of particles is $\sim 54$ per cell per species.
The physical size of the simulation box is
$L_x \times L_y = 160\lambdae \times 40\lambdae$
and therefore the size of a cell is $\Delta x = \Delta y \sim 0.039\lambdae$.
The bulk velocity of the upstream plasma is $V = 0.01c \sim 3000$km s$^{-1}$.
The temperatures of the electrons and the 
ions are $\Te = 11.3$ keV and
$\Ti = 1.25$ keV, respectively.
Thus,
the temperature ratio is given by $\Te/\Ti = 9$.
The corresponding thermal velocities are $\athe = 0.21c \sim 6.3 \times 10^4$ km s$^{-1}$ for the electrons
and $\athi = 1.6 \times 10^{-3}c$ for the ions.
The ion-acoustic speed is $c_\mathrm{s} = (\kB \Te/\mi)^{1/2} = 3.5 \times 10^{-3}c$,
where $\kB$ is the Boltzmann constant,
and the initial bulk Mach number is given by $M_\mathrm{bulk} \equiv V / c_\mathrm{s} = 2.9$.
The Debye lengths of the electrons and of the ions are $\lambda_\mathrm{De} = 0.15 \lambdae$
and $\lambda_\mathrm{Di} = 0.05 \lambdae$, respectively.
Initially, the electric and magnetic fields are zero over the simulation box.
The boundary conditions for both the particles and the electromagnetic field
are periodic in the $y$-direction.

In the simulation,
a collisionless shock is driven
according to the injection method, or the piston method.
There are two walls at the left-hand side (smaller $x$)
and the right-hand side (larger $x$) of the simulation box
and these walls reflect particles specularly.
Initially, both the electrons and the ions are loaded uniformly
in the region between the two walls
with a bulk velocity of $V$ in the $+x$-direction.
The temperatures of the electrons and the ions are equal
in the upstream.
At the early stage of the simulation,
particles that were located near the right wall
were reflected by the wall
and then interact with the incoming particles,
i.e., the upstream plasma.
This interaction causes some instability
and eventually leads to the formation of a collisionless shock.
Note that the frame of the simulation
is the downstream rest frame;
we observe the propagation of the shock
from the right to the left in the downstream rest frame.

%%%
\subsection{Results}
\begin{figure}[!hb]
\resizebox{8.cm}{!}{\includegraphics*{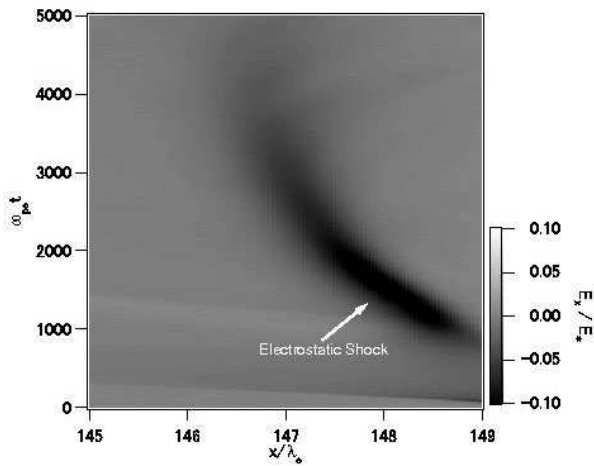}} \\[-0.6mm]
\caption{
Time evolution of the $x$-component of the electric field, $E_x$,
averaged over the $y$-direction (gray-scale).
The horizontal and vertical axes are the $x$-axis and the time,
respectively.
The electrostatic shock forms and propagates during $1000 < \omegape t < 2000$.
However,
after that time, the shock slows down and finally fades away.
}
\label{fig:Ex_dev}
\end{figure}

%%%
Figure \ref{fig:Ex_dev} shows the time evolution of the $x$-component of the electric field, $E_x$,
averaged over the $y$-direction.
We confirm that the electrostatic shock surely forms at around  $\omegape t \sim 1000$
and propagates until $\omegape t \sim 2000$
at the constant velocity of $V_\mathrm{sh,d} = 1.2 \times 10^{-3}c$
measured in the downstream frame;
the shock velocity measured in the upstream frame is given by $V_\mathrm{sh} = 1.1 \times 10^{-2} c$ 
and the shock Mach number is estimated as $M_\mathrm{s} = V_\mathrm{sh}/c_\mathrm{s} = 3.2$.
%%%
%%%

\begin{figure}[!hb]
\resizebox{8.cm}{!}{\includegraphics*{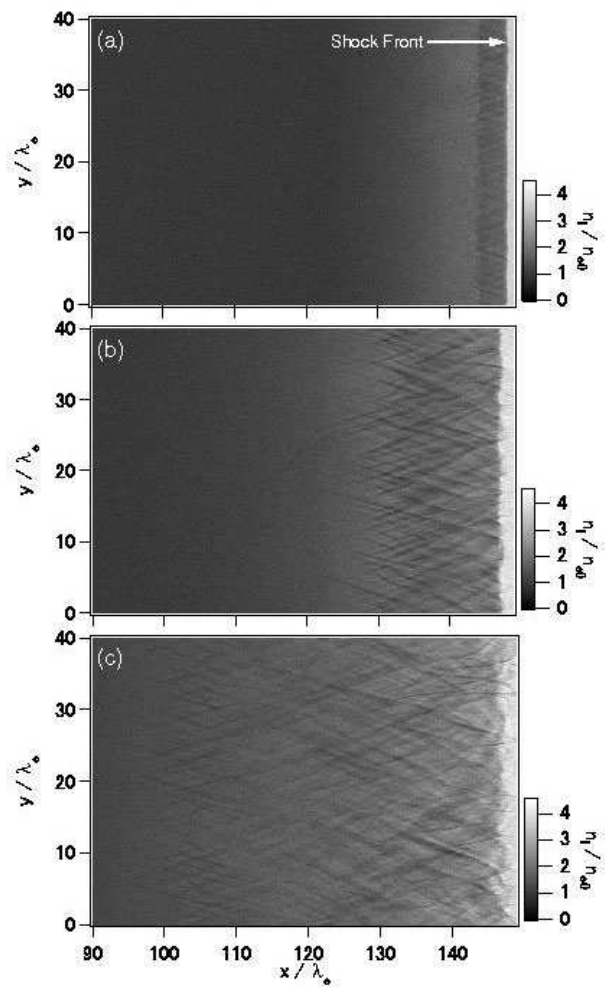}} \\[-0.6mm]
\caption{
Number density of the ions at $\omegape t =$ (a) 1500, (b) 3000, and (c) 5000.
The horizontal and vertical axes are the $x$-axis and the $y$-axis in the electron
skin depth, respectively.
The electrostatic ion-ion instability,
which generates the filamentary structure,
develops with time.
} 
\label{fig:np}
\end{figure}

\begin{figure}[!hb]
\resizebox{8.cm}{!}{\includegraphics*{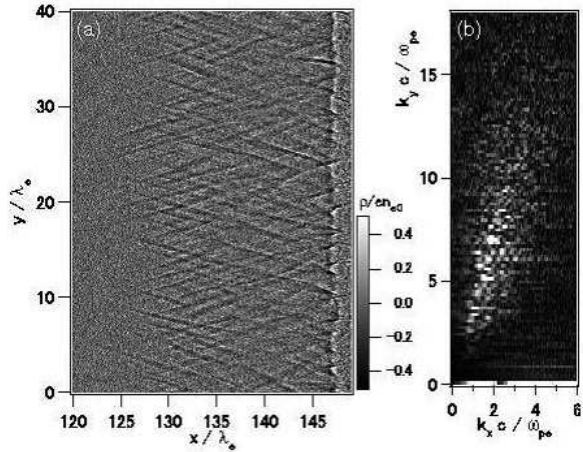}} \\[-0.6mm]
\caption{
(a) The charge density at $\omegape t = 3000$ and (b) its power spectrum.
The horizontal and vertical axes in (b) are $k_x$ and $k_y$ in $\omegape c^{-1}$, respectively.
The portion where the electrostatic mode develops shown in the panel (b)
is well agree with those obtained from the linear analysis shown in Fig.~\ref{fig:es_disp}.
}
\label{fig:rho_3000}
\end{figure}

After $\omegape t = 2000$, however, the shock slows down and finally fades away.
Figure \ref{fig:np} shows snapshots of the ion number density
at  $\omegape t = 1500$, $3000$, and $5000$.
It is evident that oblique filamentary structures develop
in front of the shock where the plasma is
under the counter-streaming condition
due to the existence of the reflected ions.
We see that with the development of the filaments,
the upstream plasma is strongly fluctuated and
the shock structure is significantly modified.
A similar structure was also observed in the simulation by \citet{Karimabadi91}.
%%%
%%%
Figure \ref{fig:rho_3000} shows the charge density at $\omegape t = 3000$ and its power spectrum,
which is directly related with the existence of the electrostatic modes.
Comparing the region where the power is large in the power spectrum
with that in the growth rate map obtained by the linear theory shown in Fig.~\ref{fig:es_disp},
it is clear that these are generated by the electrostatic ion-ion instability.
Note that, as is shown later, in fact the number density of the reflected ions is smaller than
that of the incoming ions. The ratio is typically $0.6$ [see Fig.~\ref{fig:T_ratio} (d)].
However, the growth rate as a function of $k_x$ and $k_y$ (not shown here)
is not significantly different from that of the symmetric beam case shown in Fig.~\ref{fig:es_disp};
the maximum growth rate is obtained as $\Gamma = 2.4 \times 10^{-3} \omegape$
at $|k| = 7.6\ \omegape c^{-1}$ and $\theta = 79^\circ$,
which is only slightly smaller than the symmetric case.
%%%
%%%

\begin{figure}[!hb]
\resizebox{8.cm}{!}{\includegraphics*{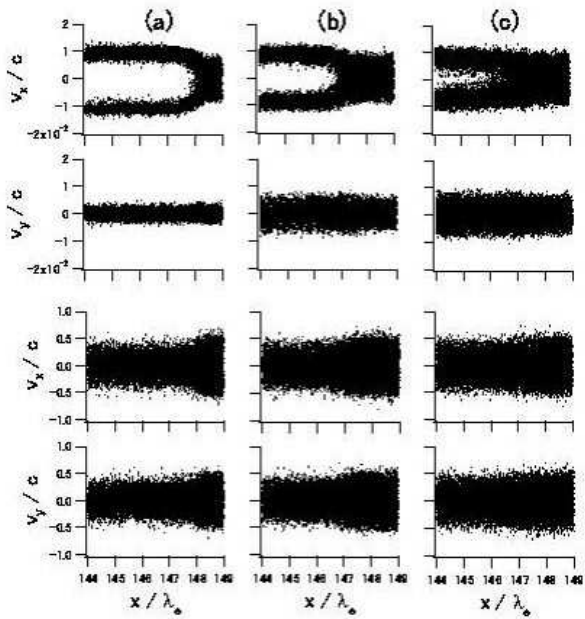}} \\[-0.6mm]
\caption{
Phase-space plots of the ions and electrons at $\omegape t =$ (a) 1500, (b) 3000, and (c) 5000:
from top to bottom,
the $x$-$v_x$ and $x$-$v_y$ distributions of the ions
and those of the electrons are shown.
In the upstream of the shock,
the ion heating is evident while the electron heating is small.
}
\label{fig:phase space}
\end{figure}

\begin{figure}[!hb]
\resizebox{8.cm}{!}{\includegraphics*{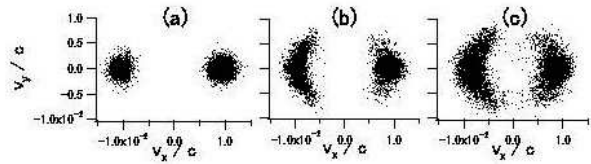}} \\[-0.6mm]
\caption{
Velocity distributions of the ions in the $v_x$-$v_y$ space
within $143 < x/\lambdae < 145$ at $\omegape t =$ (a) 1500, (b) 3000, and (c) 5000.
The heating in the $y$-direction is remarkable.
}
\label{fig:vx_vy}
\end{figure}

\begin{figure}[!hb]
\resizebox{8.cm}{!}{\includegraphics*{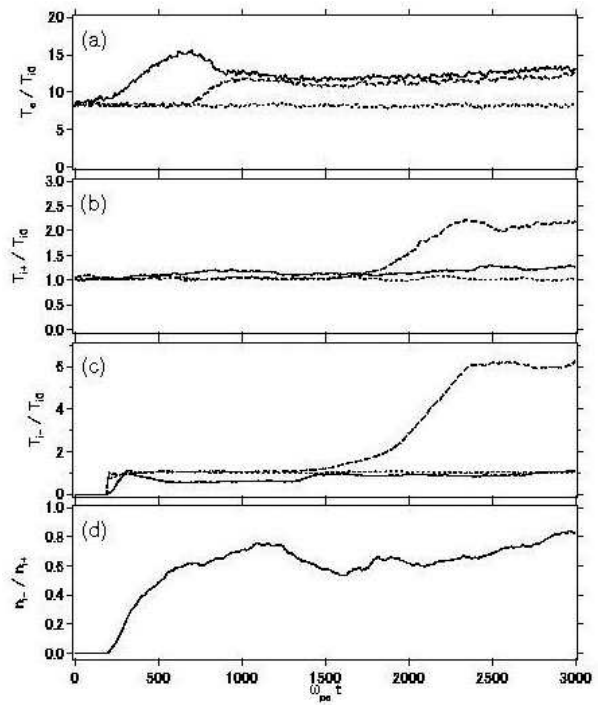}} \\[-0.6mm]
\caption{Time development of the temperature in each direction
immediately in front of the electrostatic shock
normalized to the initial ion temperature $T_{i0}$:
(a) the electrons, (b) the incoming ions, and (c) the reflected ions.
The solid curves, the dashed curves, and the dotted curves
show the $x$, $y$, and $z$ components, respectively.
It is remarkable that
the $y$ components of the temperatures of both ion populations
rapidly increase within the period $1800 < \omegape t < 2300$.
The panel (d) shows the time development of the number density ratio of the reflected ions
to the incoming ions, $n_{i-}/n_{i+}$.
}
\label{fig:T_ratio}
\end{figure}

Figure \ref{fig:phase space} shows the phase-space plots of the ions
and electrons at
$\omegape t = 1500$, $3000$, and $5000$.
The thermal velocity of the ions
in the upstream region increases with time
in both $x$ and $y$ directions,
while the increase in the velocity dispersion of the electrons is relatively small.
%%%
%%%
Figure \ref{fig:vx_vy} shows the time evolution of the velocity distributions of the ions
within $143 < x/\lambdae < 145$,
just in front of the shock front.
The ions are heated especially in the $y$ direction
because the electric field generated by the electrostatic ion-ion instability
is almost directed in the $y$-direction.
Thus,
the anisotropy in the ion temperature increases with time.
%%%
%%%

There are two possible causes for the shock decay:
one is the anisotropy in the ion temperature
(or the decrease in the electron-to-ion temperature ratio),
and the other is the fluctuation of the upstream plasma;
both are caused by the electrostatic ion-ion instability.
The time evolution of the temperatures immediately in front of the shock
are shown in Fig.~\ref{fig:T_ratio}
together with that of the ratio of the reflected ion density  to the incoming ion density.
Here, the temperatures are calculated for the particles within the region
in front of the shock given by
\begin{equation}
x_\mathrm{sh}(t) -3 \lambdae < x  < x_\mathrm{sh}(t) -1 \lambdae,
\end{equation}
where we take $x_\mathrm{sh}(t) = -V_\mathrm{sh,d} (t - 1000 \omegape^{-1}) + 148.5 \lambdae$
and $V_\mathrm{sh,d} = 1.2 \times 10^{-3} c$ from Fig.~\ref{fig:Ex_dev}.
For the ions, we distinguish the ``incoming ions'' that have $v_x > 0$
and the ``reflected ions'' that have $v_x < 0$,
while the electrons are regarded as a single population
because the mean (bulk) velocity of the electrons are negligible
compared to their thermal velocity.
Since the electrostatic shock is formed at $\omegape t \sim 1000$,
the temperatures before that time are meaningless as those in front of the shock.
We see that the electrons are heated in the $x$ and the $y$ directions in front of the shock
immediately after the shock formation and then each component of the electron temperature
keeps almost constant.
On the other hand,
the temperatures in the $y$-direction for both the incoming ions and the reflected ions
rapidly increase with time
within the period $1800 < \omegape t < 2300$,
which coincides with the period of the shock decay,
while the other components do not change significantly.
The ion heating in the $y$-direction would be caused by the electrostatic ion-ion
instability and results in the anisotropic temperatures in the ions
as well as the reduction of the electron-to-ion temperature ratio.
%%%
%%%

\begin{figure}[!hb]
\resizebox{8.cm}{!}{\includegraphics*{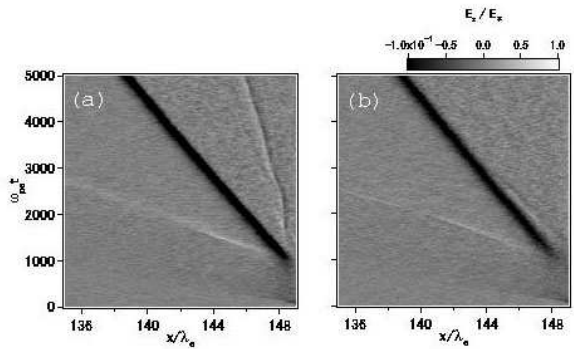}} \\[-0.6mm]
\caption{Time evolutions of the $x$-component of the electric field, $E_x$,
as in Fig.~\ref{fig:Ex_dev} obtained from quasi one-dimensional simulations (see text):
(a) for the anisotropic ion temperature
$\Tiy/\Tix=4$ and $\Te/\Tix = 9$ and (b) for the isotropic ion temperature
with $\Te/\Ti=4$.
The electrostatic shocks form and propagate without decaying in both cases.
}
\label{fig:Ex_dev_TR9_TAy4_1D}
\end{figure}

To clarify whether the anisotropic temperature in the ions
is the cause of the shock decay,
we carried out another (quasi) one-dimensional simulation
with an anisotropic ion temperature.
The simulation code used is the same two-dimensional code but we take only 8 grids in the $y$-direction
so that the simulation is essentially one-dimensional;
in such a simulation,
since the electrostatic ion-ion instability does not develop,
we can investigate the pure effect of the anisotropy in the ion temperature
on the shock formation.
For the shock formation,
the condition of the incoming ions is more important than that of the reflected ions
and Fig.~\ref{fig:T_ratio} (b) shows the typical anisotropic temperature of the incoming ions
is given by
$\Tiy \sim 2 \Tix$ around the time of the shock decay.
Here,
as a more severe case,
we take $\Tiy = 4 \Tix$ keeping the other parameters unchanged as in the two-dimensional simulation,
that is, $\Te / \Tix = 9$.
The time evolution of the electric field $E_x$ is shown in Fig.~\ref{fig:Ex_dev_TR9_TAy4_1D} (a).
We found that the electrostatic shock forms under even this condition,
where the shock speed measured in the upstream frame
is $V_\mathrm{sh} \sim 1.2 \times 10^{-2}c$ and the Mach number is given by $M_\mathrm{s}\sim 3.5$.

To confirm that the decrease in the electron-to-ion temperature ratio with isotropic ion temperature
is also not the cause of the shock decay,
we performed another quasi one-dimensional simulation with
an isotropic ion temperature but the lower electron-to-ion temperature ratio of $\Te/\Ti = 4$.
The result is shown in Fig.~\ref{fig:Ex_dev_TR9_TAy4_1D} (b)
showing that the shock is also formed with even this temperature ratio
at almost the same shock velocity and the Mach number as
the anisotropic case.
Thus,
the anisotropic temperature in the ions nor
the decrease in the temperature ratio
are not the cause of
the shock decay observed in the two-dimensional simulation.
%%%
%%%

For completeness,
we also performed another quasi one-dimensional simulation with the same parameters
as the two-dimensional one, that is, $\Te/\Ti = 9$, except the grid number in the $y$-direction.
We found that the electrostatic shock forms and propagates without decaying in this case as well (not shown)
confirming that the shock decay is a multi-dimensional effect.

\begin{figure}[!hb]
\resizebox{8.cm}{!}{\includegraphics*{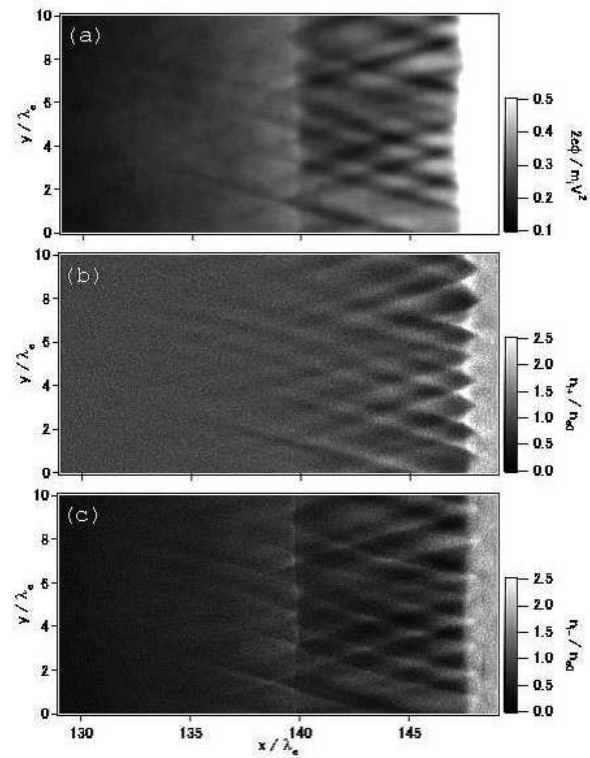}} \\[-0.6mm]
\caption{
Close-ups of
(a) The electrostatic potentials,
(b) the number density of the incoming ions ($n_\mathrm{i+}$),
and (c) that of the reflected ions ($n_\mathrm{i-}$)
at  $\omegape t = 2000$ for the two-dimensional simulation.
In (a),
the electrostatic potential in front of the shock, which is generated by the electrostatic ion-ion instability,
reaches even about a half of the shock potential.
}
\label{fig:phi_vf}
\end{figure}

\begin{figure}[!hb]
\resizebox{8.cm}{!}{\includegraphics*{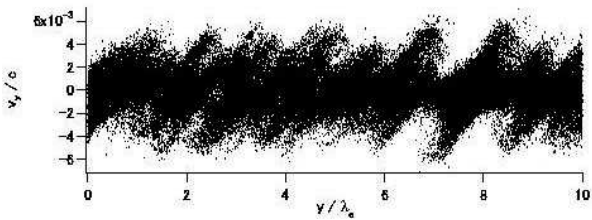}} \\[-0.6mm]
\caption{
The $y$-$v_y$ phase space distribution of the ions
within $145.5 < x/\lambdae < 145.7$
at  $\omegape t = 2000$
(c.f. Fig.~\ref{fig:phi_vf}).
The feature of the ion trapping
due to the nonlinear evolution of the electrostatic ion-ion
instability is clear.
This trapping results in the strong fluctuation in the incoming
ions as is shown in  Fig.~\ref{fig:phi_vf}(b) to destroy the shock structure.
}
\label{fig:ion_trap}
\end{figure}

Figure \ref{fig:phi_vf} shows the electrostatic potentials,
the number density of the incoming ions,
and that of the reflected ions at $\omegape t = 2000$
for the two-dimensional simulation.
The potentials are normalized by the upstream ion kinetic energy ($\mi V^2 / 2$).
It is evident that
the large potential fluctuation,
which reaches even a half of the shock potential,
exists in front of the shock due to the electrostatic ion-ion instability.
Because of this electrostatic potential,
the densities of both incoming and reflected ions are significantly disturbed
where the fluctuation pattern almost traces the waves of the electrostatic ion-ion instability.
In particular,
it can be seen that the portions where the incoming ions enter the ``shock front''
are different from those where the ions are reflected toward upstream from the ``front''.
%%%
%%%
Figure~\ref{fig:ion_trap} shows the $y$-$v_y$ phase space distribution of the ions
around $x \sim 145.5 \lambdae$.
The feature of the ion trapping \cite{Karimabadi91, Omidi88} is clear
and therefore the fluctuation of the ion densities is a result
of the nonlinear evolution of the electrostatic ion-ion instability.
Thus,
it can be concluded that
the cause of the decay of the electrostatic shock
is the large fluctuation in the ion density
due to the electrostatic ion-ion instability.
%%%
%%%

\begin{figure}[!hb]
\resizebox{8.cm}{!}{\includegraphics*{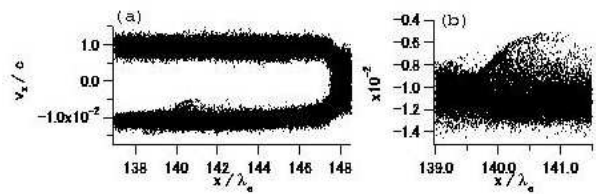}} \\[-0.6mm]
\caption{
(a) The $x$-$v_x$ phase space distribution of the ions
within $0 < y/\lambdae < 0.4$ at  $\omegape t = 2000$.
(b) The close-up of the distribution of the reflected ions around $x \sim 140 \lambdae$.
The ``secondary'' electrostatic shock exists
in the reflected ion beam around $x = 140 \lambdae$.
}
\label{fig:double_shock}
\end{figure}

In Fig.~\ref{fig:phi_vf} (c),
we can also find another interesting feature in the reflected ion density
that is seen as a discontinuity around $x \sim 140 \lambdae$,
which could also be found in Fig.~\ref{fig:np} (a).
The $x$-$v_x$ phase space distribution of the ions is
plotted in Fig.~\ref{fig:double_shock}.
As is shown in the panel (a),
there is a structure in the phase space of the reflected ions
around $x \sim 140 \lambdae$.
The panel (b) shows the close-up of this structure.
We see that some of the reflected ions are reflected again there in the opposite direction.
This structure would also be another electrostatic shock
in which the upstream region is the larger $x$ side.
This `secondary' electrostatic shock would be formed
as a result of that the faster reflected ions run into the slower ones from behind.
%%%
%%%

\begin{figure}[!hb]
\resizebox{8.cm}{!}{\includegraphics*{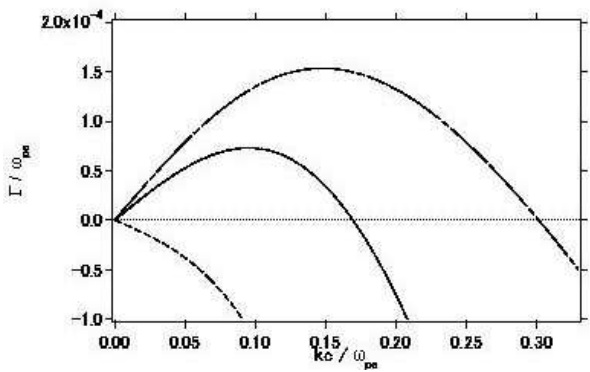}} \\[-0.6mm]
\caption{
The linear growth rates of the two Weibel-type instabilities in the anisotropic
Maxwellian beam system ($\Tiy/\Tix = 4$ and $\Te/\Tix=9$):
the beam-Weibel instability with the wave vector in the $y$-direction (solid curve)
and the (ordinary) Weibel instability with the wave vector in the $x$-direction (dashed curve).
The dot-dashed curve shows the growth rate of the beam-Weibel instability in
the isotropic Maxwellian beam system as in Sec.~\ref{sec:linear}
(i.e., $\Tiy/\Tix = 1$ and $\Te/\Ti=9$) as a reference.
The growth rate of the beam-Weibel instability is reduced due to the temperature anisotropy in the ions.
The Weibel instability in the $x$-direction is stable in this case.
}
\label{fig:em_gr_rt_aniso}
\end{figure}

As was shown in Sec.~\ref{sec:linear},
the beam-Weibel instability can develop in counter-streaming plasmas with the
wave vector in the $y$-direction as well as the electrostatic ion-ion instability.
However, the large anisotropy in the ion temperature
($\Tiy / \Tix \sim 2$ for the incoming ions
and $\Tiy / \Tix \sim 6$ for the reflected ions)
caused by the electrostatic instability
can affect the growth of the beam-Weibel instability.
In addition, this temperature anisotropy itself
may also cause the (ordinary) Weibel instability
that has the wave vector in the $x$-direction.
Here, let us consider the linear dispersion relation of these two Weibel-type instabilities
in the anisotropic Maxwellian beam system
given by the following zeroth-order distribution functions:
\begin{eqnarray}
	&&f_0^{(\mathrm{s})}(v_x, v_y, v_z) = \nonumber \\
	&&\frac{\ns}{\pi^{3/2} \athsx \athsy \athsz}
		\exp \left[
		-\frac{(v_x - \Vs)^2}{\athsx^2} - \frac{v_y^2}{\athsy^2} - \frac{v_z^2}{\athsz^2} \right],
\end{eqnarray}
where $\athsx$, $\athsy$ and $\athsz$ are the anisotropic thermal velocities for the respective directions
of the species s.
In the following, we consider the beams are symmetrical for simplicity,
that is,
$n_{e+} = n_{e-} = n_{i+} = n_{i-} = \nez$,
$V_\mathrm{e+} = V_\mathrm{i+} = V$ and $V_\mathrm{e-} = V_\mathrm{i-} = -V$,
as in Sec.~\ref{sec:linear}.
Under this condition,
the linear dispersion relation of the beam-Weibel instability with the wave vector
in the $y$-direction is given by
\begin{equation}
	\omega^2 - (kc)^2 + \sum_\mathrm{s} \omegaps^2
		\left[ \frac{\athsx^2 + 2V^2}{\athsy^2}
		\left(1 + \zetas Z(\zetas) \right) -1 \right] = 0,
\end{equation}
where $\zetas \equiv \omega/k \athsy$.
On the other hand,
the linear dispersion relation of
the (ordinary) Weibel instability with the wave vector in the $x$-direction
is given by
\begin{equation}
	\omega^2 - (kc)^2
	+ \sum_\mathrm{s} \omegaps^2
	\left[ \frac{\athsy^2}{\athsx^2}
	\left(1 + \etas Z(\etas) \right) -1 \right] = 0,
\end{equation}
where $\etas \equiv (\omega/k - \Vs)/\athsx$.
(Note that when $V=0$,
this relation is reduced to that derived by Weibel \citep{Weibel59}
without the background magnetic field.)
Figure \ref{fig:em_gr_rt_aniso} shows the growth rates of the two modes.
We take here $\Tiy/\Tix=4$, $\Tiz = \Tix$ and $\Te / \Tix = 9$
as typical values
and the other parameters are the same as those of the initial condition of the two-dimensional simulation.
As a reference, the growth rate for the isotropic beam case with $\Te/\Ti=9$
discussed in Sec.~\ref{sec:linear} is also shown in the figure.
We see that the growth rate of the beam-Weibel instability is reduced
compared with that of the isotropic beam case
due to the temperature anisotropy in the ions.
Under this condition,
the Weibel mode with the wave vector in the $x$-direction
is stable and does not grow.
(Of course, in the limit $V \to 0$, this mode should be unstable.
We confirmed that it is unstable for $V < 1.15 \times 10^{-3} c$ for the same parameters.) 
%%%
%%%

\begin{figure}[!hb]
\resizebox{8.cm}{!}{\includegraphics*{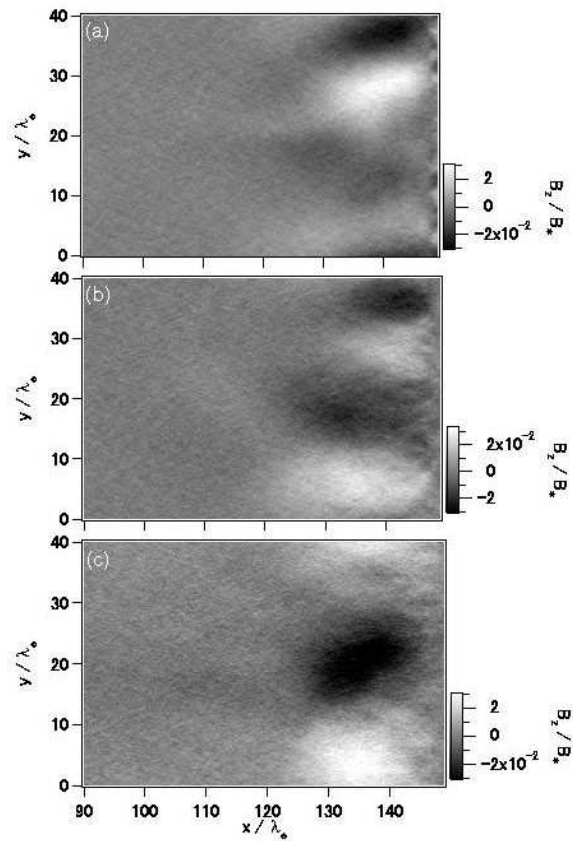}} \\[-0.6mm]
\caption{
The $z$-component of the normalized magnetic field, $B_z/B_*$, at $\omegape t =$ (a) 1500, (b) 3000, and (c) 5000
for the two-dimensional simulation,
where $B_* \equiv c (4\pi\nez \me)^{1/2}$.
The magnetic field is generated by the beam-Weibel instability in front of the shock
and it grows with time during the simulation.
} 
\label{fig:Bz}
\end{figure}

Figure \ref{fig:Bz} shows the $z$-component of the magnetic field, $B_z$,
obtained from the two-dimensional simulation
at $\omegape t=1500,\ 3000,$ and $5000$.
We see that the magnetic field grows with time with a wavelength
comparable with that derived by the linear analysis ($\sim 40 \lambdae$),
although it may be affected by the periodic boundary condition
because the wavelength is comparable with the system length in the $y$-direction.
At this time ($\omegape t \sim 5000$),
the typical magnetic field strength is $|B|/B_*\sim 2 \times 10^{-2}$
and the corresponding ion gyro-radius is given by $\sim 9\times 10^2 \lambdae$;
therefore,
the ions are hardly deflected by this magnetic field.
However, in the Weibel-type instability,
the magnetic field is generated by the current filaments
and the current filaments coalesce each other to grow larger filaments
with stronger magnetic fields
until saturation \citep[see e.g.,][]{Kato05}.
Although, because of  the limitation of the simulation box,
we cannot study the further evolution of the filaments in this simulation,
in the real world or in much larger simulations,
the magnetic field would grow much stronger with the coalescence of the filaments
and,
long after the electrostatic shock vanished,
it would become strong
enough to form the ``Weibel-mediated'' shocks,
a kind of collisionless shocks
that dissipates the upstream particle kinetic energy
via the magnetic field generated by the instability
in the shock transition region.
%%%
%%%

\begin{figure}[!hb]
\resizebox{8.cm}{!}{\includegraphics*{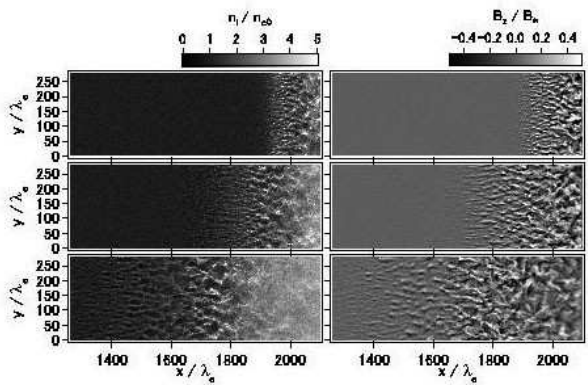}} \\[-0.6mm]
\caption{
PIC simulation of the Weibel-mediated shock
for showing the formation of the shock qualitatively
with parameters of $\mi/\me=20$ and $V=0.45c$.
The left panel shows the ion number density normalized by the upstream value, $n_i/\nez$,
and the right panel shows the $z$-component of the normalized magnetic field, $B_z/B_*$,
from top to bottom at $\omegape t = 500$, $1000$, and $2000$,
respectively.
Due to the magnetic field generated by the beam-Weibel instability,
a kind of collisionless shock,
which is called the Weibel-mediated shock,
is formed at $\omegape t \sim 2000$.
} 
\label{fig:Weibel_shock}
\end{figure}

As is reported in our previous paper of Ref.~\citep{Kato08},
we have performed a PIC simulation of the Weibel mediated shock
with the larger flow velocity ($V = 0.45c$) and the smaller mass ratio ($\mi/\me=20$)
with the equal temperatures for the electrons and the ions.
(In that simulation, the electrostatic shock and the electrostatic ion-ion instability do not develop.)
To illustrate the formation of the Weibel-mediated shock qualitatively,
the time evolutions of the ion number density and the magnetic field ($z$-component)
are shown in Fig.~\ref{fig:Weibel_shock}.
The figures show the ion current filaments develop with time due to the beam-Weibel instability
and finally the Weibel-mediated shock forms at $\omegape t = 2000$.
Note that due to the mass ratio and the flow velocity,
the temporal and spatial scales are different from those in the system
with the real mass ratio of $\mi/\me = 1836$ and the slower flow speed $\sim 3000$ km s$^{-1}$.
Returning to the situation of the present study,
the ion current filaments developing around the electrostatic shock (Fig.~\ref{fig:Bz})
would also grow further and finally will evolve into the Weibel-mediated shock
in the real world or in the simulations with much larger spatial and temporal scales.
%%%
%%%

%%%%%
\section{Conclusion}
%%%%%
We have carried out a two-dimensional electromagnetic PIC simulation
for the case of counter-streaming plasmas
at a relatively high flow velocity
with a large electron-to-ion temperature ratio of 9.
At first we have confirmed
that the electrostatic shock forms in the
early time evolution as was shown in the previous works \citep{Forslund70b,Mason72}.
We also confirmed that the electrostatic
ion-ion instability develops in front of the shock
due to the counter-streams of the ions \citep{Karimabadi91}.
Then, we found that the electric field generated by this instability
results in the strong fluctuation in the ion density and finally
leads to destroy the shock itself.
It was also found that the electromagnetic beam-Weibel instability
develops much slower than the electrostatic instability but
it becomes predominant in the later time.
This suggests the possibility that the Weibel-mediated collisionless shock is
formed according to the scenario shown in \citet{Kato08}
long after the electrostatic shock disappears.
It was also observed that the secondary electrostatic shock forms
in the reflected ions in front of the primary electrostatic shock.

\begin{acknowledgments}
We thank Y.~Ohira for helpful discussions.
We also thank Y.~Sakawa, Y.~Kuramitsu and T.~Morita
for the discussion of their experimental data.
This work was supported in part by
the Ministry of Education, Culture, Sports, Science and Technology (MEXT),
Grant-in-Aid for Young Scientists (B) (T.N.K.: 20740136).
Numerical computations were carried out at Cybermedia Center, Osaka University.
\end{acknowledgments}

%%%%%%%%%%%%
%   References
%%%%%%%%%%%%

\end{document}